\documentclass[aps,prd,twocolumn,floatfix,superscriptaddress,nofootinbib,longbibliography]{revtex4-2}

\usepackage{graphicx}
\usepackage{bm}
\usepackage{amsmath}
\usepackage{amssymb}
\usepackage{dcolumn}
\usepackage[colorlinks=true,linkcolor=blue,citecolor=blue,urlcolor=blue]{hyperref}

\newcommand{\MeV}{\,\mathrm{MeV}}
\newcommand{\fm}{\,\mathrm{fm}}
\newcommand{\Msun}{M_\odot}
\newcommand{\nB}{n_B}
\newcommand{\epsa}{\varepsilon_a}
\newcommand{\Ltil}{\tilde\Lambda}
\newcommand{\Mmax}{M_{\rm max}}
\newcommand{\gB}{g_B}
\newcommand{\load}{f_\chi m_\chi}

\begin{document}

\title{Neutron-star dark-matter limits measure the maximum-mass headroom}

\author{Probit J Kalita}
\author{Soham Singh}
\author{Pinku Routaray}
\author{Bharat Kumar}
\affiliation{Department of Physics and Astronomy, National Institute of Technology, Rourkela 769008, India}

\date{\today}

\begin{abstract}
Neutron stars constrain dark matter that is present inside them, through their
maximum mass and the tidal deformability. The standard approach adds a dark
component to one nuclear equation of state (EoS) and compares the stars with
pulsar masses and with GW170817, so each limit inherits an EoS dependence that
has not been quantified. In this paper we add a QCD axion and a $Z'$-portal
dark fermion to the relativistic mean-field (RMF) model, in the formulation of
the public code of Kumamoto et al.\ [Phys.\ Rev.\ D \textbf{112}, 043008 (2025)],
and compute the stars for $257$ published RMF parameter sets from the
compilation of Sun, Bhattiprolu, and Lattimer [Phys.\ Rev.\ C \textbf{109},
055801 (2024)], with the
dark matter either moving with the baryons or in its own equilibrium. We find
that the stars constrain only the dark rest mass per baryon, and that its upper
limit is the maximum-mass headroom of the EoS above the pulsar-mass bound
divided by a quasi-universal rate, whose residual follows the compactness of
the maximum-mass star; a compact dark core roughly doubles the rate, and the axion
condensate barely changes it. Capture cannot supply the dark matter the stars
could detect, and the portal reaches the EoS only at couplings disfavoured by
anomaly cancellation, so a neutron-star limit on dark matter is a statement
about the headroom of the nuclear EoS.
\end{abstract}

\maketitle

\section{Introduction}
\label{sec:intro}
Neutron stars test dark matter in two ways. Captured particles can heat the
star~\cite{Kouvaris2008,Baryakhtar2017,BramanteRaj2024} or make it
collapse~\cite{GoldmanNussinov1989,BramanteRaj2024}, and a dark component
already present changes its mass and radius~\cite{Ellis2018} and its tidal
deformability~\cite{Ellis2018,Nelson2019}. Studies of the second kind add dark
matter to a nuclear equation of state (EoS). Some treat it as a single fluid,
coupled to nucleons through a Higgs portal~\cite{Das2019} or produced by
neutron decay and in chemical equilibrium with the neutrons~\cite{Thakur2025};
others treat it as a second fluid coupled by gravity
alone~\cite{Ellis2018,Kain2021,Das2022,Thakur2024,Barbat2024,Karan2026}. They
confront the resulting stars with the heaviest pulsars~\cite{Antoniadis2013,
Fonseca2021} and with the tidal deformability of GW170817~\cite{Abbott2019},
and report limits on the dark fraction, the dark-matter mass and the mediator
couplings. That a dark core lowers the maximum mass is known from these
studies~\cite{Ellis2018,Kain2021,Barbat2024}. How the size of that reduction
depends on the nuclear EoS has not been established.

This dependence matters because each limit is obtained with one nuclear EoS,
and it is then unclear whether the limit describes the dark sector or that EoS.
Some EoSs fail the GW170817 bound on their own and pass once dark matter is
added: NL3 in Ref.~\cite{Das2019}; H4, MS1 and MS1B, which can no longer be
excluded with a dark core of about $5\%$ of the stellar mass~\cite{Ellis2018};
and the stiffest of the nuclear EoSs considered in
Refs.~\cite{Thakur2024,Barbat2024}. In Ref.~\cite{Thakur2024} the correlation
between the dark mass fraction and the maximum mass is strong for a single
nuclear EoS but is largely lost once four nuclear EoSs are combined. A second question is whether the operators that couple the
dark sector to baryons reach the EoS at all, and whether capture can supply the
dark component that the star would need. For weakly interacting dark matter,
capture typically deposits no more than $\sim10^{-10}\,\Msun$ over a stellar
lifetime of about $10$~Gyr (Refs.~\cite{Ellis2018,Kain2021} and references
therein), so a measurable dark component needs another origin.

In this paper we answer both questions in a model where the couplings to
baryons are explicit. The model contains a QCD axion and an asymmetric Dirac
fermion~\cite{KaplanLutyZurek2009,Zurek2014} that couples to baryons through a
vector $Z'$ of gauged baryon number~\cite{Tulin2014}. The axion potential, and
hence the axion mass, is mildly reduced at finite baryon
density~\cite{Balkin2020}. An axion lighter than the QCD value can condense in
neutron-star matter, and it condenses more readily the more the
$\sigma$ mass falls with the pion mass~\cite{Kumamoto2025}. The full calculation and its verification
are given in a companion paper~\cite{Companion}. Here we evaluate the model, in
the formulation of the public code of Ref.~\cite{Kumamoto2025}, for the $257$
usable relativistic mean-field (RMF) parameter sets compiled in
Ref.~\cite{Sun2024}. These sets have been studied before. Dutra et
al.~\cite{Dutra2014} tested the saturation properties of hundreds of RMF sets
against nuclear-matter constraints, Sun et al.~\cite{Sun2024} tabulated the
radii, tidal deformabilities and maximum masses of their stars, and Nam, Lim
and Holt~\cite{NamLimHolt2026} showed that the maximum mass follows mainly from
the saturation density, the Dirac effective mass and the $\omega$
self-coupling. We use the same sets to ask how a dark-matter limit depends on
the EoS. We show four results.
(i) For dark matter that moves with the baryons, the stellar observables
depend on the dark sector through a single number.
(ii) The limit on that number obeys a quasi-universal
relation~\cite{YagiYunes2017}: it is the maximum-mass headroom of the nuclear
EoS divided by a rate that is the same to within $11\%$ for the $19$ RMF sets
that pass the observational tests. The rate follows the compactness of the
maximum-mass star rather than the properties of nuclear matter
[Figs.~\ref{fig:headroom} and~\ref{fig:rmf}].
(iii) Capture falls short of these limits by twelve to thirteen orders of
magnitude.
(iv) The $Z'$ affects the EoS only at couplings disfavoured by anomaly
cancellation.

We make three approximations, and we state the scope of each.

First, the dark component follows the baryons, $n_\chi=f_\chi\nB$, in a single
fluid. With a portal as weak as anomaly cancellation requires
(Sec.~\ref{sec:results}), the dark matter couples to baryons essentially only
through gravity. In two-fluid treatments light dark matter then forms an
extended halo that raises the tidal deformability, while heavy dark matter
forms a compact core that lowers it~\cite{Ellis2018,Nelson2019,Barbat2024,Karan2026};
which of the two forms depends on the particle mass~\cite{Kain2021}. We
therefore also solve this gravity-only limit for all $19$ sets
[Sec.~\ref{sec:results}, Fig.~\ref{fig:twofluid}]. The limit stays proportional to the headroom while
the dark matter forms a compact core, with a rate about twice the
single-fluid one at $m_\chi=500\MeV$. A limit should therefore be quoted with
both the headroom and the assumed dark profile.

Second, we treat the axion transition in the bulk and neglect the domain-wall
tension retained in Ref.~\cite{Kumamoto2025}, which that work finds subleading
for this class of models. The effect of the transition on $\Ltil$ relies on a
positive $\sigma$-mass sensitivity to the quark mass. Without it the
condensate forms only at higher density, or not at all, and $\Ltil$ is
unchanged; the rate is unaffected either way (Sec.~\ref{sec:sectors}).

Third, we consider cold, non-rotating stars and $m_\chi\gtrsim250\MeV$, for
which the stars depend on the dark sector through the loading alone. For the
same reason the upper bounds on the non-rotating maximum mass inferred from
the remnant of GW170817~\cite{Rezzolla2018} bound the star with whatever dark
matter it carries, not the nuclear headroom, and cannot be inverted through
our relation. For comparison with two-fluid studies, which quote dark mass
fractions, our reference limit is a dark share of the rest mass of $8.1\%$, the
same at every $m_\chi$ in this regime.

We keep $\hbar c$ explicit in the equations of state and quote energy densities
and pressures in $\MeV\fm^{-3}$ and masses in solar masses. Section~\ref{sec:model}
gives the model, the parameter sets and the observational constraints,
Sec.~\ref{sec:results} the results, and Sec.~\ref{sec:concl} the conclusions.

\section{Model}
\label{sec:model}

We use one model throughout: an RMF functional of nucleons, mesons and
leptons, extended by a QCD axion and by a dark fermion coupled through a $Z'$
portal. We evaluate it for the $257$ usable meson-exchange parameter sets
tabulated in Ref.~\cite{Sun2024} (Table~\ref{tab:classes}); the parameters of
the axion and dark sectors are collected in Table~\ref{tab:params}.

\begin{table}[tb]
\caption{Parameters of the axion and dark sectors at the benchmark, and the
constants of the public code of Ref.~\cite{Kumamoto2025} that we adopt.}
\label{tab:params}
\begin{ruledtabular}
\begin{tabular}{lll}
Quantity & Symbol & Value \\
\hline
Quark-mass ratio & $z=m_u/m_d$ & $0.468$ \\
Pion mass, decay constant & $m_\pi$, $f_\pi$ & $134$, $94\MeV$ \\
Axion potential depth & $\epsa$ & $1$ \\
Nucleon sigma terms & $\sigma_N$, $\Delta\sigma$ & $50$, $1\MeV$ \\
$\sigma$-mass sensitivity & $K_\sigma$ & $0.08$ (Scenario B) \\
Dark-fermion mass & $m_\chi$ & $500\MeV$ \\
$Z'$ mass, couplings & $m_{Z'}$, $\gB=g_{\chi Z'}$ & $1$~GeV, $0.01$ \\
Neutron, proton mass & $M_n$, $M_p$ & $939.57$, $938.27\MeV$ \\
 & $\hbar c$ & $197.3\MeV\fm$ \\
\end{tabular}
\end{ruledtabular}
\end{table}

\begin{table}[tb]
\caption{The $257$ RMF parameter sets by class of functional: the number of
sets, those whose maximum mass without dark matter or condensate reaches
$2.01\,\Msun$, those that also pass the tidal test of Eq.~\eqref{eq:tests}, and
those that reach $2.01\,\Msun$ with the axion condensate.}
\label{tab:classes}
\begin{ruledtabular}
\begin{tabular}{lcccc}
Class & Sets & Reach & Pass & Reach, \\
 & & $2.01\,\Msun$ & Eq.~\eqref{eq:tests} & condensate \\
\hline
Nonlinear $\sigma\omega\rho$ & 159 & 118 & 11 & 112 \\
With cross couplings & 88 & 38 & 6 & 37 \\
Density dependent & 9 & 8 & 1 & 8 \\
With $\delta$ meson & 1 & 1 & 1 & 1 \\
\hline
All & 257 & 165 & 19 & 158 \\%
\end{tabular}
\end{ruledtabular}
\end{table}

\subsection{RMF Lagrangian with axion and $Z'$ portal}
\label{sec:sectors}
We start from the mean-field energy density of nucleons, mesons and the
axion~\cite{SerotWalecka1986,Sun2024},
\begin{equation}
\begin{aligned}
\mathcal E_B={}&\sum_{i=n,p}\mathcal E_{{\rm kin},i}(k_{Fi},M_i^*)
+g_\omega\omega n+\tfrac12g_\rho\rho\,(n_n-n_p)\\
&+V(\sigma)-F(\omega,\sigma)-G(\rho,\omega,\sigma)
+\frac{m_\delta^2\delta^2}{2(\hbar c)^3}+V_a(\theta),
\end{aligned}
\label{eq:EB}
\end{equation}
where $n=n_n+n_p\equiv\nB$ is the baryon density, $\mathcal E_{{\rm kin},i}$
is the kinetic energy density of a Fermi gas of nucleons with Fermi momentum
$k_{Fi}$ and Dirac mass $M^*_{n,p}=M_{n,p}(\theta)-g_\sigma\sigma\mp
g_\delta\delta$, the meson fields are in MeV, and
\begin{equation}
\begin{aligned}
V&=\frac{\sigma^2}{(\hbar c)^3}\Big[\tfrac12m_\sigma^2(\theta)
+\sigma\Big(\frac A3+\frac B4\sigma\Big)\Big],\\
F&=\frac{\omega^2}{(\hbar c)^3}\Big[\tfrac12m_\omega^2
+g_\omega^2\Big(g_\sigma\sigma\big[\alpha_1+\tfrac12\alpha_1'g_\sigma\sigma\big]
+\tfrac C4g_\omega^2\omega^2\Big)\Big],\\
G&=\frac{\rho^2}{(\hbar c)^3}\Big[\tfrac12m_\rho^2
+g_\rho^2\Big(g_\sigma\sigma\big[\alpha_2+\tfrac12\alpha_2'g_\sigma\sigma\big]
+\tfrac12\alpha_3'g_\omega^2\omega^2\Big)\Big],
\end{aligned}
\label{eq:VFG}
\end{equation}
where $m_\delta$ in Eq.~\eqref{eq:EB} also depends on $\theta$.
Equation~\eqref{eq:VFG} contains every term of the constant-coupling sets we
use: the nonlinear $\sigma$ and $\omega$ self-couplings, and the
$\sigma$--$\omega$, $\sigma$--$\rho$~\cite{Furnstahl1997} and
$\omega$--$\rho$~\cite{HorowitzPiekarewicz2001,ToddRutelPiekarewicz2005} cross
couplings, in the notation of Refs.~\cite{Sun2024,Dutra2014}. The $\delta$
meson enters only through $M^*_{n,p}$~\cite{Bunta2003,Zhao2024}. In the
density-dependent sets the nonlinear terms vanish, and the couplings depend
instead on $u=n/n_0$ [Eq.~(67) of Ref.~\cite{Sun2024}],
\begin{equation}
\begin{aligned}
\Gamma_j(n)&=\Gamma_j(n_0)\,a_j\frac{1+b_j(u+d_j)^2}{1+c_j(u+d_j)^2}\quad(j=\sigma,\omega),\\
\Gamma_\rho(n)&=\Gamma_\rho(n_0)\,e^{-a_\rho(u-1)},
\end{aligned}
\label{eq:DD}
\end{equation}
This density dependence adds a rearrangement term to the chemical potentials,
\begin{equation}
\Sigma_R=-\Gamma'_\sigma\sigma n_s+\Gamma'_\omega\omega n
+\tfrac12\Gamma'_\rho\rho(n_n-n_p),
\label{eq:SigmaR}
\end{equation}
where $n_s$ is the scalar density and primes denote $d/dn$.

The axion field $\theta=a/f_a$ enters in two ways. The first is the
leading-order chiral potential~\cite{DiVecchiaVeneziano1980,GrilliDiCortona2016},
which we normalise as in Eq.~(5) of Ref.~\cite{Kumamoto2025},
\begin{equation}
V_a=-\frac{\epsa m_\pi^2f_\pi^2}{(\hbar c)^3}\Big[\sqrt{1-b\sin^2(\theta/2)}-1\Big],
\label{eq:Va}
\end{equation}
with $b=4z/(1+z)^2$. We use the values of the public code of
Ref.~\cite{Kumamoto2025}: $z=m_u/m_d=2.2/4.7=0.468$, which lies within the
$0.48(3)$ of Ref.~\cite{GrilliDiCortona2016}, and $m_\pi=134\MeV$ and $f_\pi=94\MeV$
(Table~\ref{tab:params}). The second is the $\theta$ dependence of the hadron
masses. In the parametrisation of Eqs.~(9) and~(21) of Ref.~\cite{Kumamoto2025},
with $f(\theta)=[1-b\sin^2(\theta/2)]^{1/2}$,
\begin{equation}
\begin{aligned}
M_{n,p}(\theta)&=M+\sigma_N(f-1)\mp\Delta\sigma\,\frac{f-1}{f},\\
m_{\sigma,\delta}(\theta)&=m_{\sigma,\delta}\left[\frac{1+d\,f}{1+d}\right]^{1/2},
\qquad d=\frac{2K_\sigma}{1-2K_\sigma},
\end{aligned}
\label{eq:thetamass}
\end{equation}
where $\sigma_N=50\MeV$ and $\Delta\sigma=1\MeV$ are the isoscalar and
isovector nucleon sigma terms, and $K_\sigma=d\ln m_\sigma/d\ln m_q$ measures how
fast the $\sigma$ mass falls with the quark mass. Reference~\cite{Kumamoto2025}
varies only $m_\sigma$, keeps $m_\omega$ and $m_\rho$ fixed and has no $\delta$
meson. We give the $\delta$ meson the same sensitivity; this is our
assumption. The untuned QCD axion has $\epsa=1$. We take $K_\sigma=0.08$,
Scenario~B of Ref.~\cite{Kumamoto2025}; with our $z$ it gives $5.2\%$ for the
reduction of $m_\sigma$ at $\theta=\pi$.

The dark sector is
\begin{align}
\mathcal L\supset\;&\bar\chi\big(i\gamma^\mu\partial_\mu-m_\chi
-g_{\chi Z'}\gamma^\mu Z'_\mu\big)\chi
\nonumber\\
&-\tfrac{\gB}{3}Z'_\mu\sum_q\bar q\gamma^\mu q
-\tfrac{g_{aZ'Z'}}{4f_a}aZ'_{\mu\nu}\tilde Z'^{\mu\nu},
\label{eq:Ldark}
\end{align}
Equation~\eqref{eq:Ldark} is a vector-portal model. The dark fermion carries
the charge of a gauge boson of baryon number~\cite{CaroneMurayama1995,Tulin2014},
which makes it naturally stable~\cite{Tulin2014}, and its relic density is
asymmetric~\cite{KaplanLutyZurek2009,Zurek2014}. The same structure, a
spin-$\frac12$ asymmetric fermion whose vector mediator carries baryon number,
is the dark-matter model of Ref.~\cite{Nelson2019}, used in later neutron-star
inference~\cite{Rutherford2025}. Vector mediators between fermionic dark
matter and nucleons have also been added to RMF descriptions of neutron
stars~\cite{GuhaSen2021,GuhaSen2024,Kumar2026VP,Issifu2026}. The
$g_{\chi Z'}$ term alone is the vector self-repulsion of fermionic dark matter
in two-fluid studies~\cite{Nelson2019,Rutherford2025,Grippa2026}. The last
term of Eq.~\eqref{eq:Ldark} couples the axion to the hidden gauge field, as
in Refs.~\cite{Kitajima2018,Agrawal2020}; we show below that it does not reach
the EoS.

The dark matter is the fermion $\chi$. The $Z'$ itself cannot be the dark
matter, because at $m_{Z'}\simeq1$~GeV it decays to hadrons~\cite{Tulin2014}.
Spin-1 (Proca) dark matter in neutron stars~\cite{Jockel2024} is a different
model, which we do not treat.

We hold the dark density at $n_\chi=f_\chi\nB$, so $f_\chi$ is the number of
dark fermions per baryon. This single-fluid, fixed-fraction prescription is
that of Refs.~\cite{Issifu2026,IssifuMirror2026}, which assume that the dark
matter has thermalised and moves with the baryons. It replaces the fixed dark
Fermi momentum of earlier single-fluid studies~\cite{Das2019,Kumar2026VP},
which is thermodynamically inconsistent~\cite{Hajkarim2025}. The single fluid
is one of two limits of the dark distribution: (1) in the single fluid the
dark matter is tied to the baryons and moves with them; (2) a dark number
conserved separately from baryon number settles into its own equilibrium,
$\mu_\chi e^{\nu}={\rm const}$~\cite{Hajkarim2025}, which for a portal as weak
as anomaly cancellation allows is the gravity-only two-fluid star. We compute
both limits (Sec.~\ref{sec:results}) and quote the rate for each.

We take $m_\chi=500\MeV$, $m_{Z'}=1$~GeV and $\gB=g_{\chi Z'}=0.01$ unless
stated otherwise (Table~\ref{tab:params}). These values are an EoS benchmark,
not a cosmological one. Their ratio, $\gB/m_{Z'}=10^{-5}\MeV^{-1}$, lies below
the anomaly-cancellation expectation, and the $Z'$ changes no result below. We
discuss the relic abundance in Sec.~\ref{sec:results} for $m_{Z'}<m_\chi$,
where the symmetric component can annihilate into $Z'$ pairs. The benchmark
$m_{Z'}=2m_\chi$ sits on the $s$-channel resonance of $\chi\bar\chi$
annihilation; this matters for the relic density but not for the cold,
degenerate matter considered here. As in Refs.~\cite{Kumar2026VP,Issifu2026},
the three quarks give the nucleon the coupling $\gB$, and the $Z'$ mean field
\begin{equation}
Z_0'=\frac{\gB\nB+g_{\chi Z'}n_\chi}{m_{Z'}^2}
\label{eq:Z0}
\end{equation}
shifts both nucleon chemical potentials by $\gB Z_0'$. This common shift leaves
the $\beta$-equilibrium condition and the meson fields unchanged and adds
$\frac12m_{Z'}^2Z_0'^2$ to the energy density; $\chi$ adds the Fermi gas of
Eq.~\eqref{eq:fermigas} below.

Which operators reach the EoS follows from symmetry. In static, homogeneous,
spin-saturated matter the pseudoscalar density
$\langle\bar\psi i\gamma^5\psi\rangle$ vanishes at $\theta=0$ and $\pi$, where
the theory is $CP$ invariant, so local axion--nucleon couplings carry no mean
field there; the derivative coupling vanishes with the gradient and with the
spin density. The anomalous term is proportional to
$\mathbf E_{Z'}\!\cdot\!\mathbf B_{Z'}$, and static matter has no dark current
to source $\mathbf B_{Z'}$. What survives is the $\theta$ dependence of the
hadron masses. It lowers the curvature of $V_a$ in proportion to the scalar
density and drives a first-order transition to $\langle\theta\rangle\simeq\pi$.

We take matter to be cold, charge neutral and in $\beta$ equilibrium,
\begin{equation}
\begin{aligned}
&\mu_n=\mu_p+\mu_e,\qquad \mu_\mu=\mu_e,\qquad n_p=n_e+n_\mu,\\
&\mu_{n,p}=E_{F,n,p}+g_\omega\omega\pm\tfrac12g_\rho\rho+\gB Z_0'.
\end{aligned}
\label{eq:beta}
\end{equation}
The total energy density $\mathcal E$ adds the lepton, $\chi$ and $Z'$
contributions to $\mathcal E_B$. The pressure is $P=\sum_i\mu_in_i-\mathcal E$,
and the meson fields make $\mathcal E$ stationary,
$\partial\mathcal E/\partial(\sigma,\omega,\rho,\delta)=0$.

We solve these equations as the public code of Ref.~\cite{Kumamoto2025}
does, with its constants (Table~\ref{tab:params}) and its form of the meson
potential, that of
Ref.~\cite{Furnstahl1997}, onto which Eq.~\eqref{eq:VFG} maps term by term.
We add the $\delta$ meson and the density-dependent couplings in the same
form. For the parameter sets contained in that code, our equations reproduce
its energy density and pressure to numerical precision. The one exception is its
QMC-RMF2 entry, which takes the $\sigma$ and $\omega$ couplings of QMC-RMF1 in
the nonlinear terms; we use those of Ref.~\cite{Alford2022}.

As in Ref.~\cite{Kumamoto2025}, $\theta$ takes only the values $0$ and $\pi$.
At each baryon chemical potential the phase with the higher pressure is
realised, so the transition is first order, with a density jump at constant
pressure. Because $\chi$ couples to nothing but the $Z'$, the dark sector adds
the same function of $\nB$ to both phases. We therefore compute the phases
once per parameter set and repeat the construction for each dark loading.

For the crust we take the tabulated EoS of the same code, labelled there as
that of Ref.~\cite{BPS1971}. We join it to the core by the same rule: the
crust is used wherever its pressure at the same baryon chemical potential is
the higher. With dark matter we keep the join pressure of the
dark-matter-free star, so the crust carries no dark matter. This omission
changes the limits by about $1\%$ at most, because less than $1\%$ of the
baryons of the maximum-mass stars of QMC-RMF2--4 lie below the join.

We obtain the stars from the Tolman--Oppenheimer--Volkoff (TOV) equations and
the tidal deformability from the routine of the same code. That routine
integrates the equation for the static quadrupolar perturbation of
Ref.~\cite{Hinderer2010}, with $c_s^{-2}$ taken from a finite difference of the
tabulated EoS and with no matching condition at a density jump. We make two
changes. First, the routine receives a coarsely sampled stellar profile, which
misplaces a density jump at low pressure and can even return a negative
$\Ltil$. We sample the profile $40$ times more finely. $\Ltil$ then agrees to
within $0.5\%$ with the standard integration with the matching condition of
Ref.~\cite{Postnikov2010}, except for two light sets, BM-B and EMFT2, whose
crust is replaced by uniform matter almost to the surface.
Second, the conversion between pressure and solar masses in the code is
$4.9\%$ larger than the value from the CODATA 2018 constants~\cite{CODATA2018}
and the nominal solar mass parameter of IAU 2015 Resolution
B3~\cite{Prsa2016}. We use these values throughout; the correction lowers
every mass by about $2.3\%$.

Our reference set is QMC-RMF3. At the transition its density jumps by about a
third, and the neutron Dirac mass falls by about a quarter. The binary
tidal deformability $\Ltil$ of GW170817 falls from $439$ to $293$, whereas the
maximum mass hardly moves, from $2.138$ to $2.147\,\Msun$, because the
condensed layer lies far below the densities of the maximum-mass core. The
transition relies on the $\sigma$-mass shift ($K_\sigma>0$) and lies above
saturation. It should be noted that this transition is an extrapolation of
the parametrisation of Ref.~\cite{Kumamoto2025}, not a prediction. That work
estimates condensation below $n_0$ at this $K_\sigma$ for $\epsa\lesssim0.8$,
notes that its axion--nucleon model breaks down at high density, and leaves
the fate of matter near saturation to more detailed study.

To test how much rests on $K_\sigma$, we repeated the calculation for the $19$
sets that pass the constraints of Sec.~\ref{sec:obs} with $K_\sigma=0$ and a
fixed $\delta$ mass. This is Scenario~A of Ref.~\cite{Kumamoto2025}, in which
only the nucleon masses depend on $\theta$. Without the $\sigma$-mass shift
the condensate forms only at high density, above $3.4\,n_0$, and in eight
of the $19$ sets not at all inside the maximum-mass star. The $1.365\,\Msun$
stars do not reach it, so $\Ltil$ is unchanged. No set drops below
$2.01\,\Msun$, and the rate $\mathcal R$ of Eq.~\eqref{eq:Lmaxdef} changes by
less than $1\%$. The effect of the condensate on $\Ltil$ therefore
requires $K_\sigma>0$; the headroom relation does not.

Our benchmark lies on the QCD-axion line, $\epsa=1$, in Scenario~B
($K_\sigma=0.08$) of Ref.~\cite{Kumamoto2025}, whose $\varepsilon$ is our
$\epsa$. There the transition lies in the core, above saturation density, and
the crust is that of normal matter. The benchmark is therefore outside the
region of the plane of axion mass and decay constant that crust thermal
relaxation, the cooling of isolated stars and pulsar glitches exclude, which
in this scenario reaches $\epsa\simeq0.4$ [Ref.~\cite{Kumamoto2025}, Table~VII
and Fig.~18]. Treated in the bulk, the transition does not depend on $m_a$ or
$f_a$ separately, so our results add no bound in that plane. The depth of the
potential matters only through the headroom. With the code of
Ref.~\cite{Companion}, raising $\epsa$ from $0.75$ to $2$ changes the limit of
QMC-RMF3 by about $30\%$, through its headroom, while its rate changes by about
$1\%$. For lighter axions the transition moves
below saturation density, at $\epsa\lesssim0.75$ in this scenario, and into the
crust at $\epsa\lesssim0.56$ [Ref.~\cite{Kumamoto2025}, Table~III]. There the
domain wall between the two phases must be resolved. We have not tested the
headroom relation in that regime, although the relation is set by the core of
the maximum-mass star.

We take the parameters from Tables~V--VII of Ref.~\cite{Sun2024}, which
compiles $270$ RMF and point-coupling parameter sets and tabulates their saturation properties and, for stars with
pure-neutron-matter cores, their radii, tidal deformabilities and maximum
masses; the set names follow that reference. We use the sets as published and
refer to the original papers for their construction. The meson-exchange sets
come from
Refs.~\cite{Agrawal2010,Alford2022,Bender1999,Bunta2003,Cai2012,Centelles1992,Centelles1998,ChenPiekarewicz2015,Chung2000,Dadi2010,Dhiman2007,Estal2001,Fattoyev2010,Fattoyev2010a,Fattoyev2020,Furnstahl1997,Ghosh1995,Glendenning1982,Glendenning1991,GlendenningBook,Gmuca1992,Gmuca1992a,Haddad1999,Haidari2008,Horowitz2002,HorowitzPiekarewicz2001,Klahn2006,Kumar2006,Kumar2018,Lalazissis1997,Lalazissis2005,Lalazissis2009,Li2023,Liu2002,Long2004,Muller1995,MullerSerot1996,NerloPomorska2004,Niksic2002,Piekarewicz2002,Piekarewicz2006,Piekarewicz2009,Pradhan2023,Rashdan2001,Reinhard1986,Reinhard1988,Reinhard1989,Rufa1988,Serot1992,Serot1997,Sharma2000,Sharma2008,Shen2020,Sugahara1994,Sulaksono2005,Sulaksono2006,Taninah2020,ToddRutelPiekarewicz2005,Tolos2016,Typel1999,Typel2005,Typel2010}, except LBF and NL06--NL075, which Ref.~\cite{Sun2024} lists as a
private communication; for IUFSU, FSUGarnet, IOPB-I, DDME-X and NL1 we cite the
papers that introduced
them~\cite{Fattoyev2010,ChenPiekarewicz2015,Kumar2018,Taninah2020,Reinhard1986},
and for G2$^*$, NLB and NLC, TM1e and MPE also the original
papers~\cite{Sulaksono2005b,FurnstahlPriceWalker1987,Bao2014,Typel2018}. We
keep the $257$ meson-exchange sets whose saturation properties we reproduce and
whose solution extends through the maximum-mass star
(Table~\ref{tab:classes}); we do not use the point-coupling forces.

\subsection{Observational constraints}
\label{sec:obs}
We require of every parameter set, with no dark sector and no condensate,
that
\begin{equation}
\Mmax\ge2.01\,\Msun,\qquad \Ltil\le720.
\label{eq:tests}
\end{equation}
The first condition is the $68\%$ lower edge of the mass of PSR
J0740+6620~\cite{Fonseca2021}. The second is the upper edge of the $90\%$
highest-posterior-density interval of the low-spin GW170817
posterior~\cite{Abbott2019}, with $\Ltil$ evaluated at equal masses,
$1.365+1.365\,\Msun$. Of the $257$ sets, $165$ reach $2.01\,\Msun$ and $19$ of these also pass the tidal test
(Tables~\ref{tab:classes} and~\ref{tab:six}). We apply the tests before any
dark matter is added because a set that fails one cannot anchor a limit: the
dark loading would then be used to repair the nuclear EoS. For IOPB-I, for
example, the analysis returns an apparent lower limit on the loading, which
reflects only that its bare $\Ltil$ exceeds $720$. For the same reason we
apply the tests without the condensate, which softens the EoS near $2n_0$ and
lowers $\Ltil$ [Fig.~\ref{fig:rmf}(b)].

\subsection{Dark loading, limit and rate}
\label{sec:loading}
Heavy dark matter mainly adds rest mass. For a dark Fermi momentum
$k_F^\chi=(3\pi^2n_\chi)^{1/3}$ small compared with $m_\chi$, the dark gas is
non-relativistic,
\begin{equation}
\begin{aligned}
\varepsilon_\chi&=m_\chi n_\chi\Big[1+\tfrac{3}{10}x^2+\mathcal O(x^4)\Big],\\
P_\chi&=\tfrac15\,m_\chi n_\chi x^2\big[1+\mathcal O(x^2)\big],
\end{aligned}
\label{eq:darkgas}
\end{equation}
with $x=k_F^\chi/m_\chi$. To leading order it adds rest mass and no pressure,
\begin{equation}
\varepsilon(P)\;\to\;\varepsilon(P)+\load\,\nB(P),
\label{eq:inert}
\end{equation}
so the dark matter just makes every baryon heavier by $\load$. The stars then
depend on $f_\chi$ and $m_\chi$ only through this loading, the dark rest mass
per baryon (in MeV). We write $\Mmax(\load)$ for the maximum mass of the
loaded stars, $\Mmax^{(0)}=\Mmax(0)$ for that of the nuclear EoS alone, and
$\Delta M=\Mmax^{(0)}-2.01\,\Msun$ for its headroom above the mass test of
Eq.~\eqref{eq:tests}. We define the limit on the loading, and the rate at
which $\Mmax$ falls per unit loading averaged up to the limit, by
\begin{equation}
\Mmax\big((\load)_{\max}\big)=2.01\,\Msun,\qquad
\mathcal R=\frac{\Delta M}{(\load)_{\max}} .
\label{eq:Lmaxdef}
\end{equation}

Equation~\eqref{eq:inert} only explains why the loading is the right variable;
every limit below comes from the full calculation. With no scalar coupling,
$\chi$ enters the energy density as the free spin-$\frac12$ Fermi gas at
$n_\chi=(\load/m_\chi)\,\nB$,
\begin{equation}
\begin{aligned}
\varepsilon_\chi&=\frac{p_F E_F(2p_F^2+m_\chi^2)-m_\chi^4\ln[(p_F+E_F)/m_\chi]}
{8\pi^2(\hbar c)^3},\\
P_\chi&=n_\chi E_F-\varepsilon_\chi,
\end{aligned}
\label{eq:fermigas}
\end{equation}
with $p_F=\hbar c\,(3\pi^2n_\chi)^{1/3}$ and $E_F=(p_F^2+m_\chi^2)^{1/2}$,
plus the $Z'$ field energy. The expansion of Eq.~\eqref{eq:darkgas} is not
small everywhere. At the limit, $x$ at the centre of the maximum-mass star
lies between about $0.3$ and $0.9$, so the dark gas there is mildly
relativistic. Its pressure is still small against
that of the nucleons, so the dark component acts mainly through its rest
mass, as found for heavy dark fermions in Ref.~\cite{Issifu2026}. Because
Eq.~\eqref{eq:inert} omits $P_\chi$, it would give limits lower by
at most about $3\%$.

For every set we compute $\Mmax(\load)$ with the full Lagrangian at
$m_\chi=500\MeV$, with and without the condensate, and locate the limit to
$0.02\MeV$. The rate is therefore less precise in the six cases with
$\Delta M<0.01\,\Msun$; leaving them out does not change the largest
deviations from Eq.~\eqref{eq:headroom} quoted in Sec.~\ref{sec:results}.

\section{Results and Discussion}
\label{sec:results}

\begin{figure}[tbp]
\includegraphics[width=\columnwidth]{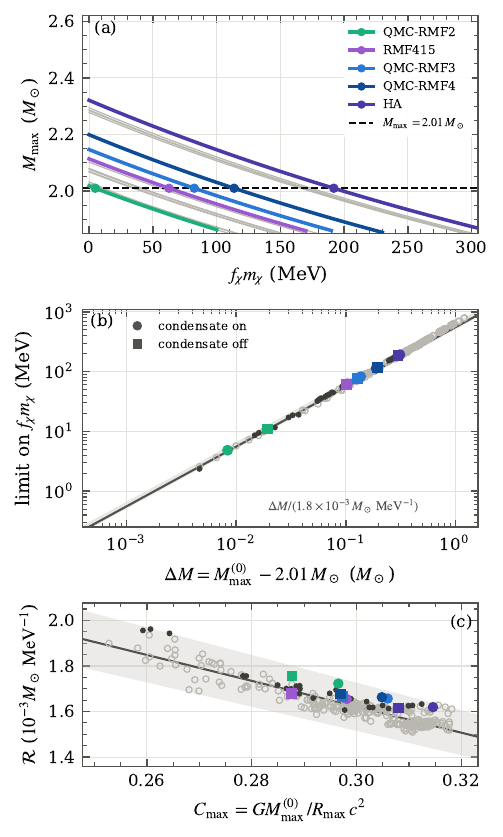}
\caption{The limit is the headroom. (a) $\Mmax$ against the dark rest-mass
loading with the axion condensate at $\epsa=1$ for the $18$ sets of
Table~\ref{tab:six} that reach $2.01\,\Msun$ with the condensate; five of them
are coloured as examples, dots mark where these cross $2.01\,\Msun$, and the
dashed line is the mass test of Eq.~\eqref{eq:tests}.
(b) The limit against the headroom $\Delta M=\Mmax^{(0)}-2.01\,\Msun$ for all
sets that reach $2.01\,\Msun$, with and without the condensate ($323$ cases):
filled if the set passes Eq.~\eqref{eq:tests}, open if it fails the tidal
test; the five examples are coloured, circles with and squares without the
condensate. The line is Eq.~\eqref{eq:headroom} and the band spans all rates.
(c) The rate $\Delta M/(\load)_{\max}$ against the compactness of the
maximum-mass star at zero loading for the same cases and symbols; the line is
Eq.~\eqref{eq:rateC} and the shaded band its largest deviation, $\pm6.5\%$.}
\label{fig:headroom}
\end{figure}

\begin{figure*}[tbp]
\includegraphics[width=\textwidth]{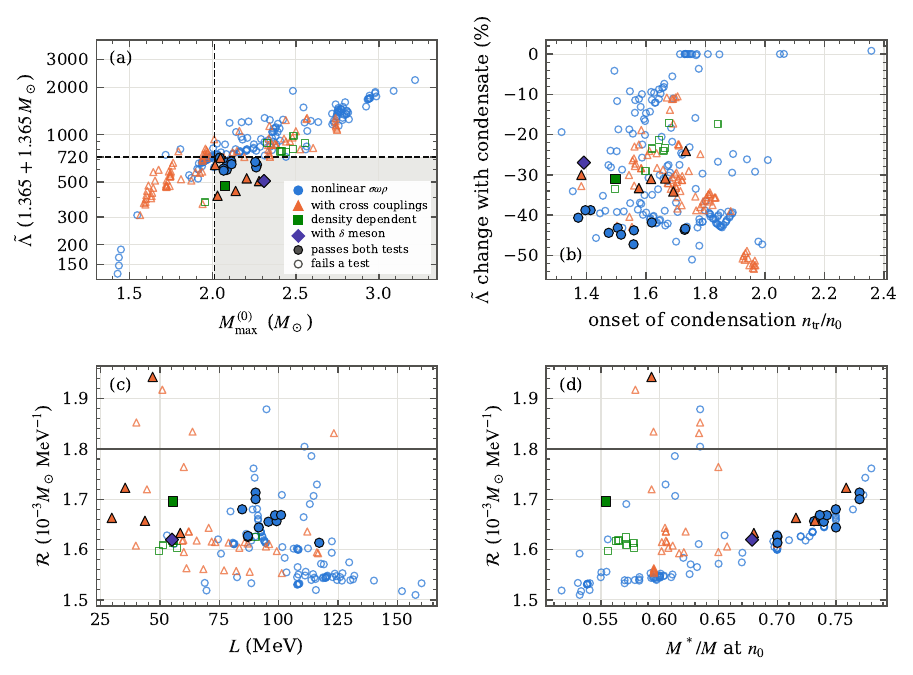}
\caption{The $257$ RMF parameter sets in $\beta$ equilibrium with the axion
and $Z'$ sectors; filled symbols pass Eq.~\eqref{eq:tests}, open symbols do
not, and marker shape and colour give the class of functional: nonlinear
$\sigma\omega\rho$ (blue circles), with cross couplings (orange triangles),
density dependent (green squares) and with a $\delta$ meson (violet diamonds). (a) $\Ltil$ at
$1.365+1.365\,\Msun$ against $\Mmax^{(0)}$, without the condensate; dashed
lines are the two tests of Eq.~\eqref{eq:tests} and the shaded corner passes
both. (b) The change of $\Ltil$ when the condensate is included, against the
density at which condensation begins, for all $257$ sets. (c), (d) The rate
$\mathcal R$ with the condensate for the $158$ sets that then reach
$2.01\,\Msun$, against the slope $L$ of the symmetry energy and against the
Dirac effective mass at saturation. The horizontal line is the rate of
Eq.~\eqref{eq:headroom}.}
\label{fig:rmf}
\end{figure*}

\begin{figure*}[tbp]
\includegraphics[width=\textwidth]{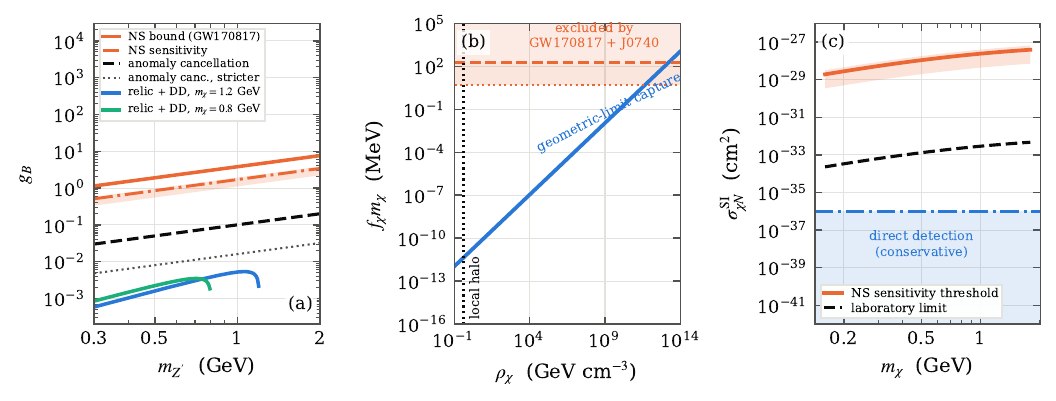}
\caption{The dark sector against laboratory and cosmological data. (a) The
$(m_{Z'},\gB)$ plane, with $g_{\chi Z'}=\gB$. Orange: the GW170817 limit of
the most constraining set that passes both tests, FSU2R (solid), and the sensitivity
threshold at which $\Mmax$ moves by $0.01\,\Msun$ (dash-dotted, reference
set; band, the $19$ sets of Table~\ref{tab:six}). Black: the
anomaly-cancellation limits of Ref.~\cite{Tulin2014} (dashed) and the
stricter reading of Ref.~\cite{DobrescuFrugiuele2014} (dotted). Blue and
green: the relic-abundance plus direct-detection limit of
Eq.~\eqref{eq:chain} for $m_\chi=1.2$ and $0.8$ GeV, drawn for
$m_{Z'}<m_\chi$. (b) The loading accreted in $10^{10}$~yr by capture at the
geometric limit against the ambient dark-matter density; dotted vertical, the
local value; orange band, the limits of Table~\ref{tab:six} with the
condensate, from $4.8$ (dotted) to $192\MeV$ (dashed), with the region above
excluded. (c) The spin-independent cross section at $m_{Z'}=1$ GeV implied by
the sensitivity threshold (orange, band as in panel a) and by the
anomaly-cancellation limit (dashed), against a conservative stand-in for the
sub-GeV limit of CRESST-III~\cite{CRESST2019} (dash-dotted; shaded,
allowed).}
\label{fig:joint}
\end{figure*}

\begin{figure}[tb]
\includegraphics[width=\columnwidth]{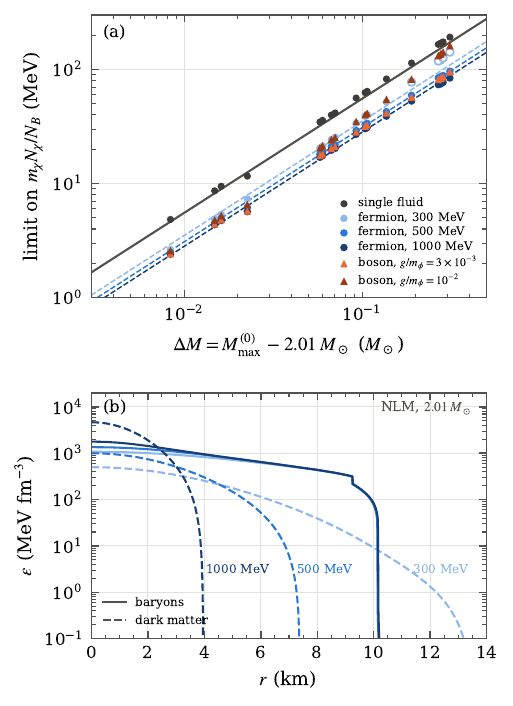}
\caption{Gravity-only two-fluid stars, with the condensate. (a) The limit on
the loading $m_\chi N_\chi/N_B$ against the headroom for the $18$ sets of
Table~\ref{tab:six} that reach $2.01\,\Msun$ with the condensate: the single
fluid (black, with Eq.~\eqref{eq:headroom}), a dark fermion at $m_\chi=300$,
$500$ and $1000\MeV$ (circles; dashed lines, the mean rate of the stars with a
dark core) and a self-repelling boson at $m_\chi=500\MeV$ (triangles). Open
symbols: the dark matter extends beyond the baryons at the limit.
(b) Energy density of the baryons (solid) and of the dark matter (dashed) in
the $2.01\,\Msun$ star of NLM at the limit, for the fermion at the three masses;
the step in the baryon profile is the condensation transition.}
\label{fig:twofluid}
\end{figure}

We first check that the stars depend on the dark sector through the loading
alone. We then give the limits for the $19$ sets that pass both tests of
Eq.~\eqref{eq:tests} (Fig.~\ref{fig:headroom} and Table~\ref{tab:six}), follow
the rate and the condensate across all $257$ sets (Fig.~\ref{fig:rmf}), solve
the gravity-only two-fluid limit (Fig.~\ref{fig:twofluid}), and compare the
limits with capture and with laboratory and cosmological bounds on the portal
(Fig.~\ref{fig:joint}).

\begin{table}[tb]
\caption{The $19$ parameter sets that pass Eq.~\eqref{eq:tests}, in order of
$\Mmax^{(0)}$ without the condensate.
$\Mmax^{(0)}$ (in $\Msun$) and $\Ltil^{(0)}$ at zero loading, the upper limit
on $\load$ (in MeV) at $m_\chi=500\MeV$, and the rate
$\Delta M/(\load)_{\max}$ (in $10^{-3}\,\Msun\MeV^{-1}$), without and with
the axion condensate at $\epsa=1$; a dash marks a set that falls below
$2.01\,\Msun$ with the condensate.}
\label{tab:six}
\begin{ruledtabular}
\begin{tabular}{lcccccccc}
& \multicolumn{4}{c}{no condensate} & \multicolumn{4}{c}{condensate} \\
Set & $\Mmax^{(0)}$ & $\Ltil^{(0)}$ & limit & rate
& $\Mmax^{(0)}$ & $\Ltil^{(0)}$ & limit & rate \\
\hline
PCSB2 & 2.015 & 642 & 2.4 & 1.96 & 2.005 & 444 & -- & -- \\
QMC-RMF2 & 2.029 & 410 & 11.0 & 1.76 & 2.018 & 270 & 4.8 & 1.72 \\
RMF422 & 2.040 & 708 & 17.2 & 1.76 & 2.025 & 399 & 8.6 & 1.71 \\
RMF423 & 2.043 & 715 & 18.6 & 1.76 & 2.026 & 405 & 9.5 & 1.70 \\
FSU2R & 2.047 & 713 & 19.0 & 1.96 & 2.033 & 542 & 11.6 & 1.94 \\
RMF419 & 2.065 & 609 & 32.3 & 1.70 & 2.068 & 336 & 34.6 & 1.67 \\
RMF421 & 2.066 & 591 & 32.9 & 1.70 & 2.070 & 329 & 35.8 & 1.67 \\
TW-99 & 2.075 & 472 & 37.8 & 1.71 & 2.080 & 326 & 41.5 & 1.70 \\
GL2 & 2.085 & 597 & 45.0 & 1.66 & 2.102 & 315 & 56.2 & 1.64 \\
NL$\rho$ & 2.087 & 667 & 44.6 & 1.72 & 2.077 & 388 & 39.9 & 1.68 \\
RMF415 & 2.112 & 676 & 60.8 & 1.68 & 2.114 & 380 & 62.7 & 1.66 \\
RMF418 & 2.114 & 650 & 61.7 & 1.68 & 2.116 & 369 & 64.3 & 1.66 \\
QMC-RMF3 & 2.138 & 439 & 76.9 & 1.67 & 2.147 & 293 & 82.5 & 1.66 \\
QMC-RMF4 & 2.205 & 528 & 116.0 & 1.68 & 2.199 & 365 & 113.7 & 1.66 \\
NLM6 & 2.256 & 671 & 153.2 & 1.61 & 2.278 & 398 & 165.9 & 1.61 \\
NLM2 & 2.262 & 620 & 155.6 & 1.62 & 2.285 & 380 & 168.8 & 1.63 \\
NLM & 2.267 & 644 & 158.6 & 1.62 & 2.287 & 394 & 170.3 & 1.63 \\
HC & 2.279 & 508 & 165.4 & 1.63 & 2.294 & 355 & 174.2 & 1.63 \\
HA & 2.309 & 508 & 185.1 & 1.61 & 2.321 & 371 & 191.9 & 1.62 \\%
\end{tabular}
\end{ruledtabular}
\end{table}

We begin with the reduction of Eq.~\eqref{eq:inert}, which we test against the full
calculation with Eq.~\eqref{eq:Ldark}. We find that it holds. For the reference
set, models with $m_\chi=300$--$1200\MeV$ and $f_\chi=0.01$--$0.16$ collapse onto
a single curve in $\load$, to better than $0.1\%$ in $\Mmax$ and about $1\%$ in
$\Ltil$. The $Z'$ enters only through its field energy, as
$(\gB/m_{Z'})^2$, and is negligible wherever it is allowed
[Fig.~\ref{fig:joint}(a), below]. The reduction fails for much lighter dark
matter, because the dark gas then becomes relativistic in the core. At
$m_\chi=100\MeV$, $\Mmax$ falls but $\Ltil$ rises as $f_\chi$ grows, whereas for
$m_\chi\ge300\MeV$ the loading lowers both. For dark matter that moves with the baryons and is not
light, stellar data therefore constrain one number and cannot separate the
dark fraction from the dark mass. A dark core in its own equilibrium behaves
differently: it lowers $\Mmax$ at a rate that depends on $m_\chi$, as we show
below with Fig.~\ref{fig:twofluid}.

Since the stars depend on the dark sector only through the loading, the next
question is how fast the maximum mass falls as the loading grows. In
Fig.~\ref{fig:headroom}(a) we show how the loading lowers the maximum mass. We see that each curve starts at $\Mmax^{(0)}$ of its set and falls
almost linearly, with a slope that decreases slightly as the loading grows;
for this reason Eq.~\eqref{eq:Lmaxdef} defines $\mathcal R$ as the average
slope up to the crossing. The curves are close to parallel, so where they
cross $2.01\,\Msun$ is fixed mainly by where they start: the five coloured
examples cross in the order of their starting masses. Only the
mass test binds. At $m_\chi=500\MeV$ the loading only lowers $\Ltil$, so for a
set that passes Eq.~\eqref{eq:tests} the tidal test never binds, and the limit
is one-sided and set by $\Mmax$ alone. For QMC-RMF3 it is $\load<82.5\MeV$ with
the condensate and $\load<76.9\MeV$ without.

Two further data leave the limits unchanged. The first is the binary mass
ratio of GW170817. At $90\%$ its low-spin posterior allows
$m_1\in(1.36,1.60)\,\Msun$~\cite{Abbott2019}. For the reference set this changes $\Ltil$ by only
a few per cent, and the limits, which are set by $\Mmax$, do not depend on the
split. The second is the NICER radius of
PSR~J0740+6620~\cite{Riley2021,Salmi2024}, which we did not include in
Eq.~\eqref{eq:tests}; at $2.07\,\Msun$ it excludes $R<11.15$~km at
$95\%$~\cite{Salmi2024}. Only two of our sets, TW-99 and GL2, fall below
this bound. Adding the radius would remove them, but it would leave unchanged
both the weakest limit and the headroom relation, which holds set by set.

The crossings of the curves in Fig.~\ref{fig:headroom}(a) are the limits.
Table~\ref{tab:six} lists them for all $19$ sets that pass both tests. We see
that the limit grows with $\Mmax^{(0)}$ down the table almost without
exception. With the condensate the limits run from $4.8\MeV$ (QMC-RMF2) to $191.9\MeV$ (HA), a factor of
forty; PCSB2 falls below $2.01\,\Msun$ once the condensate is included, which
the dashes mark. Where two sets start at nearly the same mass, as GL2 and
NL$\rho$ do, the rate decides which limit is lower. The only statement independent of the
choice among the sets is the weakest limit,
\begin{equation}
\load<192\MeV,
\label{eq:bound}
\end{equation}
which is a dark share of the rest mass, $\load/(\load+939\MeV)$, below $17\%$.

Comparing the two halves of the table isolates the condensate. It lowers
$\Ltil^{(0)}$ by a quarter to a half but moves $\Mmax^{(0)}$ by a few
hundredths of a solar mass at most, in either direction, because it softens
matter near $2n_0$ and not the maximum-mass core. Where the headroom is small,
even this small shift matters: for QMC-RMF2 it more than halves the limit. The rate
columns, by contrast, agree set by set: the condensate changes no rate by more
than $3\%$. In all cases of Table~\ref{tab:six}, $\mathcal R$ lies between
$1.61$ and $1.96\times10^{-3}\,\Msun\MeV^{-1}$.

Both the ordering of Table~\ref{tab:six} and the near-constant rate suggest
that the limit is fixed by the headroom alone. To test this directly, in
Fig.~\ref{fig:headroom}(b) we show the limit against the headroom $\Delta M$
for every set that reaches $2.01\,\Msun$. We see that on the
logarithmic axes all cases fall along a line of unit slope: the limit is
proportional to the headroom, which for the $19$ sets that pass both tests
spans nearly two orders of magnitude. To within $11\%$, and for functionals that otherwise
disagree,
\begin{equation}
(\load)_{\max}\simeq\frac{\Mmax^{(0)}-2.01\,\Msun}
{1.8\times10^{-3}\,\Msun\MeV^{-1}}.
\label{eq:headroom}
\end{equation}
The circle and the square of each coloured example lie on the same line,
because the condensate moves a set along the relation, by changing its
headroom, and not off it.

The relation does not come from a narrow choice of functionals. The $19$
sets come from all four classes of Table~\ref{tab:classes}: eleven with
nonlinear $\sigma\omega\rho$ couplings (GL2~\cite{GlendenningBook}, NLM, NLM2
and NLM6~\cite{Centelles1998}, NL$\rho$~\cite{Liu2002} and six RMF4xx sets,
RMF415--423~\cite{Dadi2010}), six with cross couplings (FSU2R~\cite{Tolos2016},
HC~\cite{Bunta2003}, PCSB2~\cite{Pradhan2023} and
QMC-RMF2--4~\cite{Alford2022}), the density-dependent TW-99~\cite{Typel1999},
and HA~\cite{Bunta2003}, with a $\delta$ meson. They span wide ranges of the
symmetry-energy slope, incompressibility, Dirac mass and radius. The sets fall into $9$ families, of which the six RMF4xx, the three
NLM and the three QMC-RMF sets are one each. Within a family the rate varies by
at most $6\%$, so the band is not produced by one over-represented family.

The relation also holds for sets that fail the tidal test [open symbols in
Fig.~\ref{fig:headroom}(b)]. Over all $323$ cases that reach
$2.01\,\Msun$ the headroom extends to above $1\,\Msun$, and
Eq.~\eqref{eq:headroom} holds to within $16\%$. A limit on dark matter obtained from one EoS is therefore a statement about that
EoS's maximum mass.

We check Eq.~\eqref{eq:headroom} against the independent calculation of
Ref.~\cite{Issifu2026}, which tabulates the maximum masses of NL3$\omega\rho$,
DD2 and FSU2R with a $200$ GeV fermion at $n_\chi=F_\chi\nB$, a loading
$F_\chi\times200$ GeV. Reading off the crossing of $2.01\,\Msun$ by linear
interpolation between their four tabulated loadings (for NL3$\omega\rho$, by
extrapolation beyond the largest), we find rates of $1.7$--$1.9\times10^{-3}\,\Msun\MeV^{-1}$, inside our band.

Equation~\eqref{eq:headroom} leaves a scatter of up to $11\%$. To understand
what sets it, in Fig.~\ref{fig:headroom}(c) we show the rate against the
compactness of the maximum-mass star. We see that the scatter about
Eq.~\eqref{eq:headroom} is not random. The rate falls as $\Mmax^{(0)}$ grows, and this drift is carried by one
property of the EoS: the compactness $C_{\max}=G\Mmax^{(0)}/R_{\max}c^2$ of its
maximum-mass star at zero loading. Over all $323$ cases,
\begin{equation}
\mathcal R\simeq\left(3.33-5.69\,C_{\max}\right)
\times10^{-3}\,\Msun\MeV^{-1},
\label{eq:rateC}
\end{equation}
with an rms deviation of $2.1\%$ and a largest deviation of $6.5\%$, for
$C_{\max}$ between $0.253$ and $0.318$. A more compact
maximum-mass star therefore needs a larger loading for the same headroom. The
compactness is the better variable: a linear fit in $\Mmax^{(0)}$ instead
doubles the largest deviation. The trend is not a property of one class of
functional, because the four classes fall on the same line. The coloured examples
follow it from QMC-RMF2 and RMF415, the least compact, to HA, the most compact.

In practice Eq.~\eqref{eq:headroom} is used in reverse, to turn a headroom
into a limit. Read as a quasi-universal relation, it is inverted with its own
scatter,
\begin{equation}
(\load)_{\max}=\frac{\Delta M}{1.8\times10^{-3}\,\Msun\MeV^{-1}}\,
(1\pm0.11),
\label{eq:invert}
\end{equation}
for EoSs that pass Eq.~\eqref{eq:tests}, and with $(1\pm0.16)$ for any of our
sets that reaches $2.01\,\Msun$. When $C_{\max}$ is known, Eq.~\eqref{eq:rateC}
reduces the uncertainty to $7\%$ over the whole range.

Figure~\ref{fig:headroom}(c) points to the maximum-mass star, rather than to
saturated matter, as the origin of the rate. To test this across the whole
compilation, and to see why so few sets pass both tests, in Fig.~\ref{fig:rmf}
we show these results among all $257$ sets. Panel~(a)
explains why so few sets pass both tests. We see that $\Ltil$ grows with
$\Mmax^{(0)}$. This is because both measure the stiffness of the EoS, so the
two tests of Eq.~\eqref{eq:tests} pull in opposite directions, and only the
$19$ sets in the shaded corner pass both. All four classes of
functional are represented in the corner.

The tidal test is also where the axion condensate enters. Panel~(b) shows
what the condensate does to the tidal deformability. With
$K_\sigma=0.08$ the axion condenses in every set. The transition begins between
about $1.3$ and $2.4\,n_0$, below the centre of the maximum-mass star in all
sets. As for QMC-RMF3, it acts on $\Ltil$ and hardly on $\Mmax$: $\Ltil$ falls
by up to $53\%$, while $\Mmax$ moves by less than $0.1\,\Msun$. The size of
the fall in $\Ltil$ depends less on the onset density than on whether the
$1.365\,\Msun$ stars reach the condensed phase. It is therefore smaller for
stiff sets, and the stiffest sets, at the top of the panel, are unchanged:
their $1.365\,\Msun$ stars are too dilute to condense. The condensate thus brings sets into line with the
tidal test, as dark matter does. In all, $61$ sets that fail it on their own
pass with the condensate, among them FSUGarnet, IOPB-I, BigApple,
DD2~\cite{Typel2010} and DD-ME2~\cite{Lalazissis2005}. This is why we apply
the tests of Eq.~\eqref{eq:tests} without the condensate. All of this panel
depends on $K_\sigma$: at $K_\sigma=0$ the condensate starts above $3.4\,n_0$
and leaves $\Ltil$ unchanged (Sec.~\ref{sec:sectors}).

Having seen that the condensate changes $\Ltil$ but not the rate, we return to
the rate itself. Panels~(c) and~(d) test whether nuclear-matter properties
organise the rate, as the compactness does. We find that they do not. The rate correlates only weakly with $L$, $K_0$,
$J$, $K_{\rm sym}$ and $M^*/M$, and sets with nearly equal $L$ can differ in
rate by a fifth. In panel~(d) the nonlinear sets rise with $M^*/M$, because a larger Dirac mass softens the EoS at high density and so
lowers $\Mmax^{(0)}$~\cite{Alford2022,NamLimHolt2026} and $C_{\max}$. The
cross-coupled sets do not follow this trend. The clearest case is the group of
seven sets with the highest rates (FSU2R, FSUGarnet, G1, IOPB-I, MS1, TM1 and
TM1e): they differ widely in $M^*/M$ and $L$, but all are among the ten least
compact sets. The rate is a property of the maximum-mass star, not of
saturated matter.

All the limits so far assume that the dark matter moves with the baryons. The
rate, however, also depends on where the dark matter sits.\label{sec:twofluid}
The single fluid is one limit of the dark distribution; the gravity-only
two-fluid star is the other. We solved the second limit for all $19$ sets of
Table~\ref{tab:six}. Here the dark matter sits in its own hydrostatic
equilibrium and couples to the nucleons only through gravity, and both fluids
obey the two-fluid TOV equations~\cite{Kain2021}. Each fluid is in
equilibrium in the common metric, so its redshifted chemical potential is
constant~\cite{Nelson2019,Hajkarim2025}. The log-enthalpies of the two fluids
therefore drop by the same amount from the centre outward, and we integrate in
that common drop; the condensation jump then enters as an exact
step. The nuclear EoS, crust and condensate are those of the single-fluid
stars. We drop the $Z'$ coupling between the two fluids, whose energy density
is below $10^{-6}$ of the baryon rest mass, and define the loading for the
whole star as $m_\chi N_\chi/N_B$, the dark rest mass per baryon.

We first ask whether the headroom relation survives. In
Fig.~\ref{fig:twofluid}(a) we show the two-fluid limits against the headroom,
in the same form as Fig.~\ref{fig:headroom}(b); Table~\ref{tab:twofluid} lists them. We
see that the limits again lie on lines of unit slope, below the single-fluid
line. For a dark fermion at $m_\chi=500\MeV$ the dark matter forms a core several
kilometres across, and the rate, $3.2$--$3.6\times10^{-3}\,\Msun\MeV^{-1}$, is about
twice the single-fluid rate of the same set. The rate roughly doubles because a core
concentrated at the centre lowers $\Mmax$ faster than the same mass spread
with the baryons. The same holds without the condensate.

The rate depends on the particle mass, because the mass sets where the dark
matter sits. To see this, we show in
Fig.~\ref{fig:twofluid}(b) the energy densities of the baryons and of the dark
matter in the limit stars of NLM. A heavier fermion contributes less Fermi
degeneracy pressure, so it forms a smaller core~\cite{Nelson2019,Kain2021} and
gives a larger rate. A lighter fermion
spreads further and can form a halo~\cite{Nelson2019,Kain2021}. At $300\MeV$ the dark matter of the six sets with the largest
headroom (HA, HC, NLM, NLM2, NLM6 and QMC-RMF4) extends beyond the baryons at
the limit (open symbols in Fig.~\ref{fig:twofluid}), and the rate falls.

A fermion is only one choice of dark matter. To test whether the result
depends on the source of the dark pressure, we repeated the test for
self-repelling bosonic asymmetric dark
matter~\cite{Nelson2019,Rutherford2023}: a boson of mass $m_\chi$ whose
repulsion is carried by a vector of mass $m_\phi$ and coupling $g$, with
\begin{equation}
\varepsilon_\chi=m_\chi n_\chi+\frac{(g/m_\phi)^2n_\chi^2}{2},\qquad
P_\chi=\frac{(g/m_\phi)^2n_\chi^2}{2},
\label{eq:boson}
\end{equation}
at $m_\chi=500\MeV$. The condensed bosons have no kinetic pressure, so the
repulsion alone supports the dark component~\cite{Nelson2019} and sets the size
of the core. For $g/m_\phi=3\times10^{-3}\MeV^{-1}$ the core is compact and the rate is
close to the fermion value. For $g/m_\phi=10^{-2}\MeV^{-1}$ the core grows with
the headroom, to about $9$~km, and the rate falls, so the limit grows more
slowly than $\Delta M$. We conclude that the dark profile sets the
rate. The proportionality to the headroom holds while the dark matter stays in
a compact core, and a limit should be quoted with both the headroom and the
assumed profile.

\begin{table}[tb]
\caption{Limits on the loading (in MeV) with the condensate, for the sets of
Table~\ref{tab:six} [Fig.~\ref{fig:twofluid}(a)]: the single fluid at $m_\chi=500\MeV$, and the
gravity-only two-fluid stars with $\load$ read as $m_\chi N_\chi/N_B$, for a
dark fermion at three masses and a self-repelling boson at $m_\chi=500\MeV$
and two values of $g/m_\phi$ (in $\MeV^{-1}$). $^h$: the dark matter extends
beyond the baryons at the limit. PCSB2, below $2.01\,\Msun$ with the
condensate, is omitted.}
\label{tab:twofluid}
\begin{ruledtabular}
\begin{tabular}{lcccccc}
& single & \multicolumn{3}{c}{fermion, $m_\chi$ (MeV)} & \multicolumn{2}{c}{boson, $g/m_\phi$} \\
Set & fluid & $300$ & $500$ & $1000$ & $3\times10^{-3}$ & $10^{-2}$ \\
\hline
QMC-RMF2 & 4.8 & 2.6 & 2.5 & 2.4 & 2.4 & 2.6 \\
RMF422 & 8.6 & 4.9 & 4.5 & 4.4 & 4.4 & 4.8 \\
RMF423 & 9.5 & 5.3 & 5.0 & 4.7 & 4.8 & 5.2 \\
FSU2R & 11.6 & 7.3 & 6.2 & 5.7 & 5.8 & 6.6 \\
RMF419 & 34.6 & 20.5 & 18.1 & 17.0 & 17.6 & 20.6 \\
RMF421 & 35.8 & 21.3 & 18.7 & 17.6 & 18.2 & 21.4 \\
TW-99 & 41.5 & 25.5 & 21.9 & 20.3 & 21.2 & 25.4 \\
GL2 & 56.2 & 33.8 & 29.1 & 27.0 & 28.4 & 35.0 \\
NL$\rho$ & 39.9 & 24.3 & 21.1 & 19.6 & 20.4 & 24.3 \\
RMF415 & 62.7 & 39.1 & 32.9 & 30.2 & 31.9 & 40.1 \\
RMF418 & 64.3 & 40.1 & 33.7 & 31.0 & 32.7 & 41.3 \\
QMC-RMF3 & 82.5 & 52.0 & 42.7 & 39.0 & 41.6 & 54.8 \\
QMC-RMF4 & 113.7 & 77.6$^h$ & 59.3 & 52.9 & 57.3 & 82.3 \\
NLM6 & 165.9 & 118.6$^h$ & 84.5 & 73.6 & 82.0 & 132.5 \\
NLM2 & 168.8 & 120.8$^h$ & 85.8 & 74.6 & 83.4 & 136.0 \\
NLM & 170.3 & 123.1$^h$ & 86.6 & 75.2 & 84.1 & 137.9 \\
HC & 174.2 & 126.4$^h$ & 88.5 & 76.9 & 86.2 & 142.5 \\
HA & 191.9 & 142.0$^h$ & 96.6 & 84.1 & 95.1 & 162.1 \\%
\end{tabular}
\end{ruledtabular}
\end{table}

Figures~\ref{fig:headroom} and~\ref{fig:twofluid} tell us how much dark
matter the stars can hold. They do not tell us whether such a loading can
arise, or whether the portal itself can be seen; these are the two questions
of Sec.~\ref{sec:intro}. In Fig.~\ref{fig:joint} we compare the limits with
capture and with laboratory and cosmological bounds on the portal.
Equation~\eqref{eq:bound} concerns percent-level dark fractions, and the
obvious source of such a
fraction is capture from the Galactic halo. Capture has a maximum that does
not depend on the cross section. Once every particle whose trajectory crosses
the star is captured, with full gravitational focusing, the accretion rate is
\begin{equation}
\dot M_{\max}=\pi R^2\rho_\chi\,\frac{v_{\rm esc}^2}{\bar v},
\label{eq:Mdot}
\end{equation}
and no larger coupling can raise it, because the star is already opaque.
Residence in a binary raises the flux by a factor of a
few~\cite{Takatsy2026}. For our $1.4\,\Msun$, $11.38$ km reference star
($v_{\rm esc}=0.60\,c$), the local density
$\rho_\chi=0.4\,{\rm GeV\,cm^{-3}}$~\cite{deSalas2021} and
$\bar v=230\,{\rm km\,s^{-1}}$, $10^{10}$~yr of accretion deliver
$6.5\times10^{-15}\,\Msun$, a loading
\begin{equation}
\load\big|_{\rm capture}\le4.4\times10^{-12}\MeV,
\label{eq:capture}
\end{equation}
which is $4.4\times10^{13}$ times below Eq.~\eqref{eq:bound}.

The local density is not the only possibility, since dark matter may be much
denser, for example in a spike~\cite{Takatsy2026}. In Fig.~\ref{fig:joint}(b) we
therefore show the same comparison at any ambient density.
We see that the accreted loading grows linearly with $\rho_\chi$ and reaches
the band of neutron-star limits only at
$\rho_\chi\sim10^{11}$--$10^{13}\,{\rm GeV\,cm^{-3}}$. Refinements do
not close this gap. Relativistic treatments~\cite{Bell2020,BramanteRaj2024}
raise Eq.~\eqref{eq:Mdot} by less than a factor of two, and nucleon
structure and strong interactions lower the capture rate~\cite{Anzuini2021}.
In a binary, dynamical friction caps the accreted mass fraction at
$\lesssim10^{-3}$ even at dark-matter-spike densities~\cite{Takatsy2026}.
Capture histories followed in time likewise change the star appreciably only
at such densities~\cite{DelPopolo2020,LiuSun2026}. A dark fraction the star can
feel must therefore have been present when the star formed or have been
produced inside it.

Production inside the star can reach the required fractions. Rough estimates
give up to a few per cent of the stellar mass, by neutron bremsstrahlung in
the newly born star or by conversion of Fermi-sea neutrons into dark particles
of nearly the neutron mass~\cite{Ellis2018} (as collected in
Ref.~\cite{Rutherford2025}). Production during the supernova~\cite{Nelson2019}
and neutron decay into dark particles~\cite{Thakur2025} are other such
channels. A dark share of a few per cent of the rest mass is a loading of tens
of MeV, inside the range of our limits, so these mechanisms, unlike capture, can reach the regime the
stars test.

Whatever its origin, the dark matter could also act on the EoS directly,
through the portal. In Fig.~\ref{fig:joint}(a) we show the portal in the
$(m_{Z'},\gB)$ plane.
In the mean field the $Z'$ enters the energy density only as
$\tfrac12m_{Z'}^2Z_0'^2\propto(\gB/m_{Z'})^2$, so the stars constrain only the
ratio, and each neutron-star curve in the panel is a straight line
$\gB\propto m_{Z'}$. At $f_\chi=0.05$ and with the condensate, $\Mmax$ of the
reference set first moves by $0.01\,\Msun$ at
$\gB/m_{Z'}\simeq1.7\times10^{-3}\MeV^{-1}$, and the other sets of
Table~\ref{tab:six} respond at similar couplings (the orange band). The strongest limit any of them
supports, from $\Ltil\le720$ for FSU2R, is
$\gB/m_{Z'}<3.8\times10^{-3}\MeV^{-1}$.

Theoretical consistency already points to far weaker couplings. Baryon number
is anomalous in the Standard Model. Cancelling the anomaly with new electroweak
fermions heavier than about $100$~GeV, which have not been seen, suggests
$\gB/m_{Z'}\lesssim10^{-4}\MeV^{-1}$~\cite{Tulin2014} (dashed). This is two
orders below the $1.4\times10^{-2}\MeV^{-1}$ at which a published EoS study
finds a visible $Z'$ effect (its set with $m_{Z'}=100\MeV$ and nucleon coupling
$1.35$)~\cite{Kumar2026VP}. For TeV-scale $Z'$ bosons the $Z'$ contribution to
the energy density is negligible~\cite{Kumar2026VP,Issifu2026}, although in
Ref.~\cite{Issifu2026} it still dominates the small baryon--dark-matter
interaction energy. A quantitative version of the anomaly argument (dotted),
together with limits on the longitudinal $Z'$, is stronger
still~\cite{DobrescuFrugiuele2014,Dror2017}. The EoS responds only a factor of
seventeen above the anomaly limit. At that limit the $Z'$ field
energy is of order $10^{-5}$ of the total energy density. The dark matter is then inert gravitating
mass, which is why the stars depend on the loading alone.

A portal strong enough to move the stars must also be visible in the
laboratory. In Fig.~\ref{fig:joint}(c) we show that direct detection closes
the remaining room. Such a portal implies, for
$g_{\chi Z'}=\gB$, $m_{Z'}=1$ GeV and $m_\chi=500\MeV$, a spin-independent
cross section
\begin{equation}
\sigma^{\rm SI}_{\chi N}=\frac{\mu_{\chi N}^2g_{\chi Z'}^2\gB^2}{\pi m_{Z'}^4}
\simeq1.0\times10^{-28}\,{\rm cm^2}.
\label{eq:sigmaSI}
\end{equation}
This is about $10^8$ times above $10^{-36}\,{\rm cm^2}$, a deliberately
conservative stand-in for the sub-GeV limit of CRESST-III, which was the most
stringent between $0.16$ and $1.8$ GeV when published~\cite{CRESST2019}. We see
that the gap persists over this whole mass range, and that even a portal at
the anomaly-cancellation limit, $1.3\times10^{-33}\,{\rm cm^2}$ at
$m_\chi=500\MeV$, lies three orders of magnitude above the stand-in. Very large
cross sections do not escape this argument. Above about
$10^{-30}\,{\rm cm^2}$ the overburden stops dark matter before it reaches a
detector deep underground such as CRESST-II. A CRESST prototype operated at
the surface, however, excludes cross sections up to about
$2\times10^{-27}\,{\rm cm^2}$ and closes the gap to the XQC limits~\cite{EmkenKouvaris2018}, so
$10^{-28}\,{\rm cm^2}$ is excluded as well. The portal is disfavoured in any
case by the anomaly argument above.

Cosmology limits the portal from the other side. Asymmetric dark matter must
have removed its symmetric component, which requires at least the thermal
annihilation cross section~\cite{Steigman2012,Planck2018,LinYuZurek2012}. In
the secluded regime $m_{Z'}<m_\chi$, where $\chi\bar\chi\to Z'Z'$
dominates~\cite{PospelovRitzVoloshin2008}, this sets a floor
$g_{\chi Z'}\ge0.018$ at $m_\chi=0.5$ GeV for $m_{Z'}\ll m_\chi$, which rises
with $m_{Z'}$. Combined with direct detection, the floor gives, for $m_\chi=0.5$--$1.2$ GeV and
$m_{Z'}=0.3$--$0.6$ GeV,
\begin{equation}
\gB/m_{Z'}<2.0\text{--}5.2\times10^{-6}\MeV^{-1},
\label{eq:chain}
\end{equation}
a further factor of $20$--$50$ below the anomaly-cancellation limit. These are
the blue and green curves of Fig.~\ref{fig:joint}(a), two to three orders of
magnitude below the neutron-star band.

\section{Conclusion}
\label{sec:concl}

We have shown that, for dark matter that moves with the baryons, neutron stars
constrain one number, the dark rest mass per baryon, and that the limit on it
is the maximum-mass headroom of the nuclear EoS divided by
$1.8\times10^{-3}\,\Msun\MeV^{-1}$. This quasi-universal relation holds to
within $11\%$ for the $19$ RMF sets that pass the mass and tidal tests, with or
without the axion condensate. These sets span nonlinear, cross-coupled,
density-dependent and $\delta$-meson functionals in $9$ families. The relation
holds to within $16\%$ for all $165$ sets that reach $2.01\,\Msun$, and its
residual is set by the compactness of the maximum-mass star, not by the
properties of saturated matter. A dark core in its own equilibrium obeys the same
proportionality while the core stays compact, with a rate about twice as
large at $m_\chi=500\MeV$ that grows with $m_\chi$.

The axion condensate changes the stars but not the relation. With
$K_\sigma=0.08$ it forms in every set and lowers $\Ltil$ by up to $53\%$; with
$K_\sigma=0$ it forms only at high density and leaves $\Ltil$ unchanged. In
both cases it changes the rate by at most $3\%$, because it affects a limit
only through the headroom.

The loading the stars test cannot come from capture, which falls short of the
limits by twelve to thirteen orders of magnitude [Fig.~\ref{fig:joint}(b)].
The portal reaches the EoS only at couplings disfavoured by anomaly
cancellation and, for $m_\chi=500\MeV$ and $m_{Z'}=1$ GeV, excluded by direct
detection [Fig.~\ref{fig:joint}(a,c)]. Neutron-star data therefore constrain
the dark-matter content of the star, whether present at its formation or
produced inside it, and not the dark-matter--nucleon coupling.

These results suggest two practices. First, a neutron-star limit on dark
matter should be quoted with the maximum-mass headroom of the EoS it rests on,
or converted through Eq.~\eqref{eq:invert}. Second, the nuclear EoS should
pass the observational tests on its own before dark matter is added; this
holds in a joint fit of nuclear and dark parameters as much as for a single
parameter set. Radius
measurements of the heaviest pulsars, which constrain $C_{\max}$, would sharpen
the relation, and the density jump at the transition would carry a
discontinuity $g$-mode~\cite{Finn1987,Sotani2001}, a possible signature that
we have not computed.

A natural next step is to test the relation systematically: (1) for rotating
and hot stars; (2) for non-relativistic and quark-matter EoSs; (3) for axions
lighter than the QCD value, whose condensate can reach the
crust~\cite{Kumamoto2025} and where the domain wall between the phases must
be resolved; and (4) for dark-matter fractions set by a capture or production
history rather than fixed in advance.

\section{Acknowledgement}
Thanks to Sasmita Mishra and H. Mishra for the insightful discussion on the Z portal.
\bibliographystyle{apsrev4-2}
\bibliography{refs}

\end{document}